# ON OPERAD STRUCTURES OF MODULI SPACES AND STRING THEORY

TAKASHI KIMURA, JIM STASHEFF, AND ALEXANDER A. VORONOV

ABSTRACT. We construct a real compactification of the moduli space of punctured rational algebraic curves and show how its geometry yields operads governing homotopy Lie algebras, gravity algebras and Batalin-Vilkovisky algebras. These algebras appeared recently in the context of string theory, and we give a simple deduction of these algebraic structures from the formal axioms of conformal field theory and string theory.

This paper started as an attempt to organize geometrically various algebraic structures discovered in 2d quantum field theory, see Witten and Zwiebach [46], Zwiebach [49], Lian and Zuckerman [29], Getzler [14, 15], Penkava and A. S. Schwarz [32], Horava [20], Getzler and J. D. S. Jones [16], Stasheff [43, 44] and Huang [21].

The physical importance of these structures is that they lead toward the classification of string theories at the tree level, because the structure constants of the algebras appear as all correlators of the theory. We suggest that an appropriate background for putting together those algebraic structures is the structure of an operad. On the one hand, as we point out, a conformal field theory at the tree level is equivalent to an algebra over the operad of Riemann spheres with punctures, cf. Huang and Lepowsky [22]. On the other hand, this one operad gives rise to several other operads creating these various algebraic structures. The relevance to physics is that theories such as conformal field theory or string-field theory provide a representation of the geometry of the moduli space of such punctured Riemann spheres in the category of differential graded vector spaces.

This paper, one of a series, deals with a part of these algebraic structures, namely with the structure of a homotopy Lie algebra and the related structures of the gravity algebra and Batalin-Vilkovisky algebra. A richer structure, the moduli space of Riemann spheres, induces a homotopy version of a Gerstenhaber algebra, which contains

Research of the first author was supported by an NSF Postdoctoral Research Fellowship
Research of the second author was supported in part by NSF grant DMS-9206929 and a Research and Study Leave from the University of North Carolina-Chapel Hill
Research of the third author was supported in part by NSF grant DMS-9108269.A03

a commutative homotopy algebra as another piece. We plan to study these structures in a subsequent paper.

The relation between homotopy Lie algebras and the moduli spaces is similar to the relation between homotopy associative algebras and the associahedra [41]. The idea of such a relation goes back to Beilinson and Ginzburg [5], Getzler [14], Getzler and Jones [16] and Ginzburg and Kapranov [17].

The main results of this paper are the construction of a real compactification of the moduli space of Riemann spheres with punctures, which is closely related to a compactification of Deligne-Knudsen-Mumford, and an explanation of the origins of the structure of a homotopy Lie algebra in a topological conformal field theory in terms of a natural stratification of this moduli space. These strata are naturally labelled by trees. We show that these strata give rise to the operad governing homotopy Lie algebras which has a nice description in the language of trees due to Hinich and Schechtman [18]. We then define a (tree level c=0) topological conformal field theory in the language of operads slightly generalizing the definitions by Segal [37, 39] and show that the strata of the real compactification of moduli space combined with certain results of minimal area metrics by Wolf and Zwiebach [46] induces the structure of a homotopy Lie algebra upon a relative subcomplex of the state space of a topological conformal field theory. We also point out that the related structures of a topological gravity and Batalin-Vilkovisky algebra can be understood in this framework, as well.

## 1. Operads

Operads were originally invented [30] for the study of iterated (based) loop spaces: for two excellent overviews of this theory, see Adams [1] and May [31]. For an updated treatment, see also Kriz and May [27]. Before that invention (and hence without the name), Stasheff created an operad [40, 42] that made explicit the higher homotopies required of the multiplication on an H-space for it to be homotopy equivalent to a loop space. He introduced a sequence of convex polyhedra $K_j$ (which have come to be known as associahedra), of dimension $j - 2$, with the property that a connected space $X$ has the homotopy type of a loop space if and only if there is a sequence of maps $\theta_j : K_j \times X^j \to X$ satisfying certain compatibility conditions.

The analogous result for $n$-fold iterated loop spaces makes use of the Boardman-Vogt little $n$-cubes operad $\mathcal{C}_n$ [7] which is a sequence of spaces $\mathcal{C}_n(j)$, $j \geq 1$, which are the spaces of imbeddings of $j$ disjoint copies of the standard cube $I^n$ in $I^n$, the embeddings being affine maps coordinatewise. The basic result is due to May [30]: a connected space admits an action of the operad $\mathcal{C}_n$ iff it has the homotopy type of an $n$-fold iterated loop space.

The case $n = \infty$ was particularly important historically; more recently, analogous operads related to Lie structures and based on various moduli spaces have become important - as will become clearer below.

The homology of topological operads gives algebraic operads; the calculations of the homology of the little n-cubes operads or of the corresponding configuration spaces due to Arnold [2] and F. Cohen [8] in the 1960–70's have recently played a key role in the physically motivated structures we are about to present.

From the utilitarian viewpoint, operads are universal objects describing various algebraic structures. The use of operads becomes essential when describing structures with multilinear operations. Before giving the definition of an operad, let us present a typical example.

**Example 1.1.** Let $X$ be a topological space. Consider the collection of sets $\mathcal{O}(n) :=$ $\mathrm{Map}(X^n, X,)$, $n \geq 1$, of continuous mappings. The following structures can be associated with this collection.

(1) The natural action of the symmetric group $\Sigma_n$ on $\mathcal{O}(n)$ by permutations of the inputs.

(2) The composition law:

$$\gamma : \mathcal{O}(k) \times \mathcal{O}(n_1) \times \cdots \times \mathcal{O}(n_k) \to \mathcal{O}(n_1 + \cdots + n_k),$$
$$(f; f_1, \ldots, f_k) \mapsto \gamma(f; f_1, \ldots, f_k) := f(f_1, \ldots, f_k).$$

(3) The unit $e := \mathrm{id}_X \in \mathcal{O}(1)$.

One can verify that the following properties are satisfied:

(4) The composition is equivariant with respect to the symmetric group actions: $\Sigma_k \times \Sigma_{n_1} \times \cdots \times \Sigma_{n_k}$ acts on the left-hand side and maps naturally to $\Sigma_{n_1 + \cdots + n_k}$, acting on the right-hand side.

(5) The composition is associative, i.e., the following diagram is commutative:

$$
\begin{array}{ccc}
\left\{ \begin{array}{c} \mathcal{O}(k) \times \mathcal{O}(n_1) \times \cdots \times \mathcal{O}(n_k) \\[4pt] \times \mathcal{O}(m_{11}) \times \cdots \times \mathcal{O}(m_{k,n_k}) \end{array} \right\} & \xrightarrow{\mathrm{id} \times \gamma^k} & \mathcal{O}(k) \times \mathcal{O}(m_1) \times \cdots \times \mathcal{O}(m_k) \\[12pt]
{\scriptstyle \gamma \times \mathrm{id}} \downarrow & & \downarrow {\scriptstyle \gamma} \\[8pt]
\mathcal{O}(n) \times \mathcal{O}(m_{11}) \times \cdots \times \mathcal{O}(m_{k,n_k}) & \xrightarrow{\quad \gamma \quad} & \mathcal{O}(m)
\end{array}
,
$$

where $m_i = \sum_j m_{ij}$, $n = \sum_i n_i$ and $m = \sum_i m_i$.

(6) The unit $e$ satisfies natural properties with respect to the composition: $\gamma(e; f) = f$ and $\gamma(f; e, \ldots, e) = f$ for each $f \in \mathcal{O}(k)$.

**Definition 1.1.** An *operad* is a collection of sets (topological spaces, vector spaces, complexes, . . . , objects of a symmetric monoidal category) $\mathcal{O}(n)$, $n \geq 1$, with data (1)–(3) (all arrows being morphisms) satisfying axioms (4)–(6). We will also call operads of complexes *differential graded (DG) operads*.

The notion of a *morphism of operads* can be introduced naturally.

*Remark* 1.1. We will also deal with operads $\mathcal{O}(n)$, $n \geq 2$, without a unit. In any case, a unit element can be formally added, so that we will have an operad in the sense of the definition above.

Here is an important linearization of Example 1.1.

**Example 1.2 (Endomorphism operad).** Let $V$ be a vector space (or a complex of vector spaces). Let

$$\mathcal{E}nd\,(V)\,(n) = \mathrm{Hom}(V^n, V), \qquad n \geq 1.$$

In analogy to the above operad of map, this is an operad of vector spaces (complexes, respectively). It is the appropriate setting for studying the composition of multilinear maps. The symmetric group will always act by permuting the inputs. In the case of complexes, this will assume the basic super (or graded) commutativity sign convention: $ab = (-1)^{|a||b|}ba$, where $|a|$ and $|b|$ are the degrees of symbols $a$ and $b$.

**Definition 1.2.** An *algebra over an operad* $\mathcal{O}$ of vector spaces (complexes) is a vector space (complex, respectively) $V$ provided with a morphism of operads:

$$(1.1) \qquad\qquad \mathcal{O}(n) \to \mathcal{E}nd\,(V)\,(n), \qquad n \geq 1.$$

This is equivalent to a sequence of maps

$$(1.2) \qquad\qquad \mathcal{O}(n) \times V^n \to V, \qquad n \geq 1$$

satisfying certain compatibility conditions.

We sometimes will use algebras over operads in the category of sets or topological spaces. It will mean that the morphism (1.1) of operads is considered in the smallest possible category which contains terms of both operads. For instance, a differential graded algebra over a topological operad is a complex $V$ provided with a morphism (1.1) of topological operads such that the image is in the component of degree 0. Usually, $V$ is considered with some topology. If $\mathcal{O}$ is an operad of manifolds, we require that the morphism be smooth.

**Example 1.3 (Commutative operad).** Consider the operad $\mathcal{C}(n) := k$ for all $n \geq 1$ in the category of vector spaces over a field $k$. The symmetric group action is taken to be trivial and the composition is nothing but multiplication in the field $k$. Then a vector space $V$ is an algebra over $\mathcal{C}$ if and only if $V$ is a commutative algebra. Indeed, to define an algebra over $\mathcal{C}$ is the same as to define a collection for all $n \geq 2$ of linear mappings

$$V^{\otimes n} \to V,$$

$$v_1 \otimes \cdots \otimes v_n \mapsto v_1 \cdot \ldots \cdot v_n,$$

which are commutative (because the symmetric group acts trivially on $\mathcal{C}(n)$) and which are compositions of the binary mapping $v_1 \otimes v_2 \mapsto v_1 \cdot v_2$ (because for $n > 2$ the space $\mathcal{C}(n)$ is generated by operad compositions of the space $\mathcal{C}(2)$).

**Example 1.4 (Associative operad).** Take the collection of vector spaces $\mathcal{A}(n) := k[\Sigma_n]$, the group algebra of the symmetric group, $n \geq 1$. Provided with the action of symmetric groups via left multiplication and the natural composition law, they form an operad whose algebras are nothing but associative algebras.

**Example 1.5 (Lie operad).** A Lie algebra is an algebra over the operad $\mathcal{L}(n) := H_{n-2}(\mathcal{M}_{n+1}, k)$ for $n \geq 2$, $\mathcal{L}(1) := k$, where $\mathcal{M}_{n+1}$ is the moduli space of Riemann spheres with $n + 1$ punctures, a complex manifold of dimension $n - 2$. The middle dimensional homology group $H_{n-2}$ inherits an operad structure from the following operad-like structure associated with the spaces $\mathcal{M}_{n+1}$. The symmetric group permutes all punctures but the $n + 1$st one, and the composition is induced by choosing a holomorphic coordinate around each puncture and sewing the Riemann spheres as in Section 3. The action of the symmetric group on $H_{n-2}(\mathcal{M}_{n+1}, k)$ also includes the sign of permutation. This identification is essentially a result of F. Cohen [9] and of Schechtman and Varchenko [33], who expressed it in the equivalent terms of the homology of configuration spaces rather than moduli spaces, and of Beilinson and Ginzburg [4].

**Definition 1.3 (Homotopy Lie algebras).** A *homotopy Lie algebra* is a complex $V = \sum_{i \in \mathbb{Z}} V_i$ with a differential $Q$, $Q^2 = 0$, of degree 1 and a collection of $n$-ary brackets:

$$[v_1, \dots, v_n] \in V, \qquad v_1, \dots, v_n \in V, \ n \geq 2,$$

which are homogeneous of degree $3 - 2n$ and super (or graded) symmetric:

$$[v_1, \dots, v_i, v_{i+1}, \dots, v_n] = (-1)^{|v_i||v_{i+1}|}[v_1, \dots, v_{i+1}, v_i, \dots, v_n],$$

$|v|$ denoting the degree of $v \in V$, and satisfy the relations

$$(1.3) \quad Q[v_1, \dots, v_n] + \sum_{i=1}^{n} \epsilon(i)[v_1, \dots, Qv_i, \dots, v_n]$$
$$= \sum_{\substack{k+l=n+1 \\ k,l \geq 2}} \sum_{\substack{\text{unshuffles } \sigma: \\ \{1,2,\dots,n\}=I_1 \cup I_2, \\ I_1=\{i_1,\dots,i_k\},\ I_2=\{j_1,\dots,j_{l-1}\}}} \epsilon(\sigma)[[v_{i_1}, \dots, v_{i_k}], v_{j_1}, \dots, v_{j_{l-1}}],$$

where $\epsilon(i) = (-1)^{|v_1|+\cdots+|v_{i-1}|}$ is the sign picked up by taking $Q$ through $v_1, \dots, v_{i-1}$, $\epsilon(\sigma)$ is the sign picked up by the elements $v_i$ passing through the $v_j$'s during the unshuffle of $v_1, \dots, v_n$, as usual in superalgebra.

*Remark* 1.2. Here we follow the physics grading and sign conventions in our definition of a homotopy Lie algebra, [46, 49]. These are equivalent to but different from those in the existing mathematics literature, cf. Lada and Stasheff [28], in which the $n$-ary bracket has degree $2 - n$. With those mathematical conventions, homotopy Lie algebras occur naturally as deformations of Lie algebras. If $L$ is a Lie algebra and $V$ is a complex with a homotopy equivalence to the trivial complex $0 \to L \to 0$, then $V$ is naturally a homotopy Lie algebra, see Schlessinger and Stasheff [34]. Similarly, with the physics conventions, homotopy Lie algebras can occur naturally as deformations of ordinary graded Lie algebras with a bracket of degree $-1$, which are equivalent to graded Lie algebras after a shift of grading and redefining the bracket by a sign, see [49, Section 4.1]. (For topologists, the physics conventions correspond to the algebra of homotopy groups of a space with respect to Whitehead product while the math conventions correspond to the algebra of homotopy groups of a loop space with respect to Samelson product.)

According to a Hinich-Schechtman theorem [18], homotopy Lie algebras can be described as algebras over a certain tree operad, which is encoded in the topology of the moduli spaces due to Beilinson and Ginzburg [4]. We are going to describe a modification of these results in the next two sections. A beautiful extension of these ideas can be found in [17].

## 2. The homotopy Lie operad

**2.1. Trees.** Here a *tree* will mean a directed connected graph, such that each vertex has at most one incoming edge and does not have exactly one outgoing edge. Obviously, a tree has an *initial vertex*, the one with no incoming edges, and may have a number of *terminal vertices*, the ones with no outgoing edges. We assume that terminal vertices of a tree are ordered with numbers $1, 2, \ldots, n$, and we will usually consider *trees with orientations*, where an *orientation* on a tree $T$ is an orientation on the real vector space $\mathbb{R}^{\mathrm{edges}(T)} \cong \mathbb{R}^{e(T)}$, where edges$(T)$ is the set of *internal edges* of the tree $T$, i.e., all edges, except the *terminal edges* (going to terminal vertices), and $e(T)$ is the number of internal edges. (The orientations are used below to get the signs right in order to have $d^2 = 0$.)

We identify trees which are isomorphic. Equivalently, we can think of trees as trees in three-space and identify those which can be deformed (isotopically) to each other. For example, there is a unique tree with $n$ terminal vertices and one *internal* (i.e., nonterminal) *vertex*, called a *corolla*, and $\binom{n}{i}$ trees with $n$ terminal and two internal vertices and $i$ outgoing edges of the upper vertex for $2 \leq i \leq n-1$:

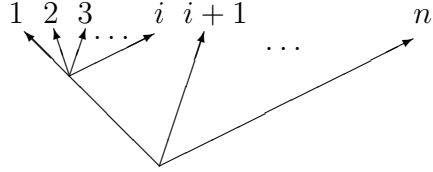

and all its *unshuffles*

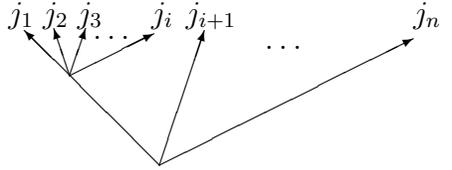

$$j_1 < \cdots < j_i, \quad j_{i+1} < \cdots < j_n,$$
$$\{j_1, \ldots, j_n\} = \{1, \ldots, n\}.$$

If orientation is taken into account, then for each nonoriented tree $T$, there are two nonequivalent oriented trees $\pm T$, corresponding to the two choices of orientation.

**2.2. Operad structure on trees.** Let $\mathcal{T}(n)$ denote the set of trees (with orientation) with $n$ terminal vertices. The symmetric group $\Sigma_n$ acts on $\mathcal{T}(n)$ by reordering the terminal vertices of a tree. For example, the symmetric group acts trivially on corollas. We can also define a *composition* of trees:

$$\gamma : \mathcal{T}(k) \times \mathcal{T}(n_1) \times \ldots \mathcal{T}(n_k) \to \mathcal{T}(n_1 + \ldots n_k).$$

It grafts given trees $T_1 \in \mathcal{T}(n_1), \ldots, T_k \in \mathcal{T}(n_k)$ to a given tree $T \in \mathcal{T}(k)$, placing the initial vertex of each $T_i$ on top of the $i$th terminal vertex of $T$. The orderings of the terminal vertices of the $T_i$'s are inserted into the ordering of the terminal vertices of $T$, and the orientation on $\mathbb{R}^{\mathrm{edges}(\gamma(T;T_1,\ldots,T_k))}$ is produced via the natural isomorphism

$$\mathbb{R}^{\mathrm{edges}(T_1)} \times \cdots \times \mathbb{R}^{\mathrm{edges}(T_k)} \times \mathbb{R}^{\mathrm{edges}(T)} \times \mathbb{R}^{t(T)} \xrightarrow{\sim} \mathbb{R}^{\mathrm{edges}(\gamma(T;T_1,\ldots,T_k))},$$

where $t(T)$ is the set of terminal edges of $T$ oriented according to the enumeration of the terminal vertices.

**Proposition 2.1.** *The sets $\mathcal{T}(n)$, $n \geq 1$, form an operad $\mathcal{T}$.*

**2.3. The tree complex.** For each $n \geq 1$, let $\mathcal{S}(n)$ be the vector space generated by elements of $\mathcal{T}(n)$ with the defining relations $T + (-T) = 0$, where $(-T)$ is the tree $T$ with the opposite orientation. We grade each vector space $\mathcal{S}(n)$ by defining the degree $|T|$ of a tree $T \in \mathcal{S}(n)$ via

$$|T| := \mathrm{int}(T) + 2 - 2n,$$

where $\mathrm{int}(T)$ is the number of internal vertices of $T$. Notice that $3 - 2n \leq |T| \leq 1 - n$ for $n \geq 2$ and $|T| = 0$ for $n = 1$.

**Lemma 2.2.**

$$|T| = \sum_v (2 - \mathrm{out}(v)) + 1 - n,$$

*where the summation runs over the internal vertices $v$ of $T$ and $\mathrm{out}(v)$ is the number of outgoing edges for $v$.*

*Proof.* Use downward induction on the number of internal vertices of the tree. $\square$

Before defining a differential on $\mathcal{S}(n)$, we will define a partial order on the set of trees, $\mathcal{T}(n)$.

**Definition 2.1.** We use $T' \to T$ to indicate that $T$ is obtained from $T'$ by contracting an internal (i.e., not incoming to a terminal vertex) edge:

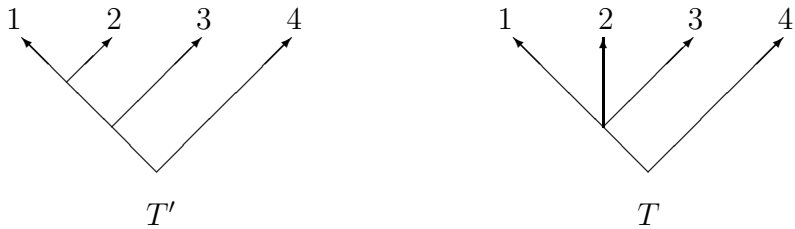

We say that $T' \leq T$, if there exists a sequence of arrows $T' \to T_1 \to \cdots \to T_k \to T$.

The contraction of an internal edge in a tree $T' \in \mathcal{T}(n)$ defines a tree $T \in \mathcal{T}(n)$ with the orientation induced from the natural embedding $\mathbb{R}^{\mathrm{edges}(T)} \hookrightarrow \mathbb{R}^{\mathrm{edges}(T')}$.

We can now define a *differential* $d : \mathcal{S}(n) \to \mathcal{S}(n)$ by the formula

$$dT := \sum_{T' \to T} T'.$$

For instance,

(2.1)

$$= \sum_{\substack{k+l=n+1 \\ k,l \geq 2}} \sum_{\substack{\text{unshuffles } \sigma: \\ \{1,2,\ldots,n\}=I_1 \cup I_2, \\ \#I_1=k, \#I_2=l-1}}$$  .

As above, an unshuffle $\sigma$ is a partition $\{1, 2, \ldots, n\} = I_1 \cup I_2$ into two ordered subsets $I_1$ and $I_2$.

Evidently, the differential $d$ can be rewritten in terms of the operad structure

$$d\delta_n = \text{sym}_n \sum_{\substack{k+l=n+1 \\ k,l \geq 2}} \binom{n}{k} \gamma(\delta_l; \delta_k, e, \ldots, e),$$

where $e := \delta_1 = \{\text{the one-vertex tree}\}$ is the unit of the tree operad $\mathcal{T}$ and $\text{sym}_n := \frac{1}{n!} \sum_{\sigma \in \Sigma_n} \sigma$ is the standard symmetrization operator.

**Proposition 2.3.**     (1) *The operator $d$ satisfies $d^2 = 0$ and $\deg d = 1$.*
   (2) *The operad structure on trees $\mathcal{T}$ induces the structure of an operad on $\mathcal{S} := \{\mathcal{S}(n) \mid n \geq 1\}$. This structure is compatible with the differential $d$, i.e., $\mathcal{S}$ is a DG operad, an operad in the category of complexes.*

**Definition 2.2.** We will call the operad $\mathcal{S} := \{\mathcal{S}(n) \mid n \geq 1\}$ the *homotopy Lie operad*.

*Remark* 2.1. The complex $\mathcal{S}(n)$ is part of Kontsevich's graph complex. The operad $\mathcal{S}$ was introduced by Hinich and Schechtman [18] as the operad generated by the corollas $\delta_n$, $n \geq 2$, with the defining relations (2.1) as an operad of graded vector spaces. This operad is the dual operad in the sense of Ginzburg-Kapranov [17] of the commutative operad $\mathcal{C}$ of Example 1.3. Beilinson and Ginzburg [4] also show that the cohomology of the operad $\mathcal{S}$ is the Lie operad $\mathcal{L}$ of Example 1.5 twisted by a certain determinantal operad (this twisting changes signs only and makes the graded Lie operad for a bracket of degree $-1$ out of the usual Lie operad), see [17].

What we need in the sequel is just the following theorem, which shows that the homotopy Lie operad $\mathcal{S}$ is a homotopy generalization of the Lie operad $\mathcal{L}$.

**Theorem 2.4.** *An algebra over the operad $\mathcal{S}$ is a homotopy Lie algebra. Each homotopy Lie algebra admits a natural structure of an algebra over $\mathcal{S}$.*

*Proof.* For a complex $V$ of vector spaces with a differential $Q$ of degree 1, $Q^2 = 0$, the structure of an algebra over the operad $\mathcal{S}$ on $V$ is a morphism of DG operads:

$$\phi : \mathcal{S}(n) \to \mathcal{E}nd\,(V)\,(n), \qquad n \geq 1,$$

where $\mathcal{E}nd\,(V)\,(n) := \operatorname{Hom}(V^{\otimes n}, V)$ is the *endomorphism operad*, which is also a DG operad (with the usual internal differential determined by $Q$). Given such a morphism $\phi$, we define the $n$-ary bracket on $V$:

$$[v_1, \ldots, v_n] := \phi(\delta_n)(v_1 \otimes \cdots \otimes v_n),$$

which is graded symmetric, because the action of the symmetric group on corollas $\delta_n$ was trivial. Note that the degree of the bracket is equal to that of the corolla $\delta_n$, which is $3 - 2n$. Since $\phi$ is a morphism of DG operads, $Q\phi = \phi d$, and in view of (2.1), this is equivalent to Equation (1.3).

Conversely, given a collection of $n$-ary brackets on $V$, $n \geq 2$, we define a morphism $\phi$ on the generators $\delta_n$ by the above formula. The operad $\mathcal{S}$ is generated by the corollas $\delta_n$ with the defining relations (2.1), see [17] and [18], so the mappings $\phi$ really define a morphism of operads if the relations (2.1) are satisfied by the $\phi(\delta_n)$'s. Equations (1.3) show that this is the case. At the same time, this means that $Q\phi = d\phi$, and thus, $\phi$ is a morphism of DG operads. $\quad\square$

## 3. Moduli spaces

### 3.1. The Deligne-Knudsen-Mumford compactification. We recall the Deligne-Knudsen-Mumford compactification of the moduli space

$$\mathcal{M}_n := ((\mathbb{CP}^1)^n \setminus \Delta)/\operatorname{PGL}(2, \mathbb{C})$$

of $n$-punctured complex projective lines $\mathbb{CP}^1$. Here $\Delta = \{\text{diagonals}\} = \{(z_1, \ldots, z_n) \in (\mathbb{CP}^1)^n \mid z_i = z_j \text{ for some } i \neq j\}$ and $n \geq 3$, see [11, 23, 24, 25] or any review on two-dimensional quantum field theory. The compactification $\overline{\mathcal{M}}_n$ is itself the moduli space (the base of a universal family) of stable $n$-punctured complex curves, which include nonsingular punctured projective lines as well as degenerations of them of a certain kind. A *stable $n$-punctured complex curve of genus* 0 is a connected compact complex curve $C$ of genus 0 with $n$ punctures, such that (1) it may have ordinary double points away from the punctures, (2) each irreducible component of the curve $C$ is a projective line and (3) the total number of punctures and double points on each component of $C$ is at least 3. Both $\mathcal{M}_n$ and $\overline{\mathcal{M}}_n$ are complex algebraic manifolds of complex dimension $n - 3$. The moduli space $\mathcal{M}_n$ of nonsingular curves is an open submanifold in the projective manifold $\overline{\mathcal{M}}_n$. The complement is a divisor, formed by all degenerate curves.

One can visualize the degeneration of a punctured projective line (a punctured Riemann sphere) as a process where the sphere undergoes "mitosis" into two spheres by forming a long thin neck away from the punctures. An equation of the neck is locally $z_1 z_2 = \epsilon$, $\epsilon \in \mathbb{C}$, and as $\epsilon \to 0$, the equation turns into $z_1 z_2 = 0$, which means that the sphere degenerates into two spheres joined at a double point. The degenerations must all be stable, that is, there must be at least three punctures or double points on each irreducible component.

A construction of the Deligne-Knudsen-Mumford compactification as the base of a universal family of stable $n$-punctured complex curves is obtained by blowing up $(\mathbb{CP}^1)^{n-3}$ or $\mathbb{CP}^{n-3}$, see [4, 23, 24, 25]. There is also a similar compactification of the configuration space $((\mathbb{CP}^1)^n \setminus \Delta)$, which is more regular and symmetric, see Fulton-MacPherson [13] and Beilinson-Ginzburg [4], which can give compactifications of the moduli space, by taking the quotient of the union of strata of dimension greater than 2 with respect to the group $\mathrm{PGL}(2, \mathbb{C})$.

The combinatorics of the Deligne-Knudsen-Mumford compactification can be easily described by trees.

**Theorem 3.1** ([4, 17, 23]). *There is a stratification of $\overline{\mathcal{M}}_{n+1}$, such that*

(1) $\overline{\mathcal{M}}_{n+1} = \coprod_{T \in \mathcal{T}(n)/\mathrm{or}} S_T$, *where $\mathcal{T}(n)/\mathrm{or}$ is the set of all trees without orientation.*

(2) *Each stratum $S_T$ is a smooth connected locally closed algebraic subvariety and* $\dim_{\mathbb{C}} S_T = 1 - n - |T|$.

(3) $S_T \subset \overline{S}_{T'} \Leftrightarrow T \leq T'$.

(4) *There is a unique open stratum $S_{\delta_n} \cong \mathcal{M}_{n+1}$.*

(5) *There is a natural isomorphism $S_T \xrightarrow{\sim} \mathcal{M}_{k+1} \times \mathcal{M}_{n_1+1} \times \cdots \times \mathcal{M}_{n_k+1}$, where the tree $T \in \mathcal{T}(n)$ is the composition of the corollas $\delta_k$, $\delta_{n_1}, \ldots, \delta_{n_k}$ in the sense of the operad structure on $\mathcal{T}$.*

(6) $\partial S_T = \sum_{T \to T'} \overline{S}_{T'}$, *where $\partial$ is the boundary.*

Here is a way of "recognizing" the tree corresponding to a punctured stable curve with double points. Take terminal vertices corresponding to the first $n$ punctures, the $n+1$st puncture will be the initial vertex. All other vertices of the tree correspond to the double points. The initial vertex is joined by an outgoing edge to all punctures and double points lying on the same irreducible component. These double points are joined by outgoing edges to punctures and other double points if they share the same irreducible component, and so on.

Given a $k+1$-punctured stable curve and $k$ stable curves with $n_1 + 1, \ldots, n_k + 1$ punctures, respectively, we can form the union of all these curves with a total of $n_1 + \cdots + n_k + 1$ punctures by attaching the $i$th curve at its $n_i + 1$st (initial) puncture to the $i$th puncture of the $k+1$-punctured curve. The enumeration of the remaining punctures is by inserting the orders, as usual. The action of the symmetric group on $\overline{\mathcal{M}}_{n+1}$ is by permutations of the first $n$ (terminal) punctures. The natural isomorphism (5) of Theorem 3.1 implies that there is a unique structure of a stable algebraic curve on the union. This construction leads to the following corollary.

**Corollary 3.2.** *The composition of stable curves described above provides $\{\overline{\mathcal{M}}_{n+1} \mid n \geq 1\}$ with the structure of an operad of algebraic varieties. This structure is compatible with the operad structure on trees via the correspondence $T \mapsto S_T$ of Theorem 3.1.*

**3.2. A real compactification of $\mathcal{M}_n$.** This compactification $\underline{\mathcal{M}}_n$ will also be a moduli space of stable punctured curves of genus 0 decorated with certain phase parameters attached to each double point. The compactified space will be a compact oriented differentiable manifold with corners, fibered over the strata of the Deligne-Knudsen-Mumford compactification with the fibre equal to $(S^1)^p$, where $p$ is the complex codimension of the Deligne-Knudsen-Mumford stratum. Thus, $\dim \underline{\mathcal{M}}_n = \dim \mathcal{M}_n = 2n - 6$. Here and henceforth all dimensions are real. The compactification we construct here is new although a similar real compactification of the configuration space, in which case it is the real version of the Fulton-MacPherson compactification, has been considered by Getzler [14] and Kontsevich [26] as well as Axelrod and Singer. [3]

A stable $n$-punctured complex curve $C$ of genus 0 is *decorated with relative phase parameters at double points* if the sum of the arguments of germs of holomorphic coordinates on each irreducible component at each double point is chosen. Geometrically, this additional data can be thought of as a choice of tangent directions (defined by making the argument equal to 0) at each double point on each irreducible component of $C$ modulo the diagonal action of rotations $S^1$ at each double point. The degeneration process can be described as the same "mitosis" into two spheres by forming a neck, but now in the local equation $z_1 z_2 = \epsilon$ of the neck, we let $|\epsilon| \to 0$, leaving the argument $\arg \epsilon$ fixed, and mark the directions $\arg z_1 = 0$ and $\arg z_2 = 0$ on the neck. This defines the two tangent directions on the irreducible components of the degeneration.

The moduli space of such objects can be constructed by making real blowups (i.e., pasting in copies of $S^1$) along the irreducible components of the divisor $D = \overline{\mathcal{M}}_n \setminus \mathcal{M}_n$ of the Deligne-Knudsen-Mumford compactification. These components are the closed strata $\overline{S}_T$ for all trees $T \in \mathcal{T}(n)$ with two internal vertices. Since the divisor $D$ is a normal crossing divisor, the result does not depend on the order of the blowups. We denote the result of these blowups by $\underline{\mathcal{M}}_n$. It is a real analytic manifold with corners, its interior is $\mathcal{M}_n$ and, therefore, $\underline{\mathcal{M}}_n$ is homotopically equivalent to $\mathcal{M}_n$.

Recall that $\overline{\mathcal{M}}_{n+1} \to \overline{\mathcal{M}}_n$ is a universal family of stable $n$-punctured curves. The natural projection $\overline{\mathcal{M}}_{n+1} \to \overline{\mathcal{M}}_n$ (forgetting the $n+1$st puncture) lifts to a morphism $\underline{\mathcal{M}}_{n+1} \to \underline{\mathcal{M}}_n$. This yields the following theorem.

**Theorem 3.3.** *The morphism $\underline{\mathcal{M}}_{n+1} \to \underline{\mathcal{M}}_n$ is a universal family of stable $n$-punctured complex curve of genus $0$ decorated with relative phase parameters at double points.*

Let us define a stratification of $\underline{\mathcal{M}}_n$ by defining each stratum on $\underline{\mathcal{M}}_n$ as the preimage of a stratum on $\overline{\mathcal{M}}_n$ via the natural projection $\underline{\mathcal{M}}_n \to \overline{\mathcal{M}}_n$. Evidently, the combinatorics of the stratification will be the same, that is, the following variant of Theorem 3.1 will hold for $\underline{\mathcal{M}}_n$.

**Theorem 3.4.** *There is a stratification of $\underline{\mathcal{M}}_{n+1}$, such that*

(1) $\underline{\mathcal{M}}_{n+1} = \coprod_{T \in \mathcal{T}(n)/\text{or}} S_T^r$.

(2) *Each stratum $S_T^r$ is a smooth connected locally closed submanifold and $\dim_{\mathbb{R}} S_T^r = -|T| - 1$.*

(3) $S_T^r \subset \overline{S}_{T'}^r \Leftrightarrow T \leq T'$.

(4) *There is a unique open stratum $S_{\delta_n}^r \cong \mathcal{M}_{n+1}$.*

(5) *There is a natural projection $S_T^r \to \mathcal{M}_{k+1} \times \mathcal{M}_{n_1+1} \times \cdots \times \mathcal{M}_{n_k+1}$ with fibre $(S^1)^k$, where the tree $T \in \mathcal{T}(n)$ is the composition of corollas $\delta_k,\ \delta_{n_1}, \ldots, \delta_{n_k}$ in the sense of the operad structure on $\mathcal{T}$.*

(6) $\partial S_T^r = \sum_{T \to T'} \overline{S}_{T'}^r$.

**3.3. Chain and homology operads.** Item (5) prevents $\{\underline{\mathcal{M}}_{n+1} \mid n \geq 2\}$ from being an operad. But some operad structure is naturally defined on the (singular, for instance) chain complexes $C_\bullet(\underline{\mathcal{M}}_{n+1})$ of $\underline{\mathcal{M}}_{n+1}$'s with coefficients in the ground field.

Define the composition as follows. Given a collection $C,\ C_1, \ldots,\ C_k$ of chains in $C_\bullet(\underline{\mathcal{M}}_{k+1}),\ C_\bullet(\underline{\mathcal{M}}_{n_1+1}), \ldots,\ C_\bullet(\underline{\mathcal{M}}_{n_k+1})$, respectively, we take their product $C \times C_1 \times \cdots \times C_k$ in $\underline{\mathcal{M}}_{k+1} \times \underline{\mathcal{M}}_{n_1+1} \times \cdots \times \underline{\mathcal{M}}_{n_k+1}$ and then the preimage (the transfer, to be more precise) $\gamma(C; C_1, \ldots, C_k)$ in the space $\overline{S}_T^r \subset \underline{\mathcal{M}}_{n_1+\cdots+n_k+1}$ with respect to the natural projection $\overline{S}_T^r \to \underline{\mathcal{M}}_{k+1} \times \underline{\mathcal{M}}_{n_1+1} \times \cdots \times \underline{\mathcal{M}}_{n_k+1}$, where $T$ is the tree $\gamma(\delta_k; \delta_{n_1}, \ldots, \delta_{n_k})$. In other words, by choosing a singular fundamental chain $S$ for the fibre $(S^1)^k$, the preimage of the chain $C \times C_1 \times \cdots \times C_k$ can be expressed as $S \times C_1 \times \cdots \times C_k$ to define $\gamma(C; C_1, \ldots, C_k)$ as an element of $C_\bullet(\underline{\mathcal{M}}_{n_1+\cdots+n_k+1})$. Thus, we obtain a composition:

$$(3.1) \quad \gamma : C_\bullet(\underline{\mathcal{M}}_{k+1}) \otimes C_\bullet(\underline{\mathcal{M}}_{n_1+1}) \otimes \cdots \otimes C_\bullet(\underline{\mathcal{M}}_{n_k+1}) \to C_\bullet(\underline{\mathcal{M}}_{n_1+\cdots+n_k+1}),$$

which is a morphism of complexes, except that it has degree k. Therefore we regrade $C_\bullet$ so that the degree of a chain $C$ is now $-\dim C - 1$. The action of the symmetric group $\Sigma_n$ on chains $C_\bullet(\underline{\mathcal{M}}_{n+1})$ comes from the natural action on $\underline{\mathcal{M}}_{n+1}$ by permutations of the first $n$ punctures.

**Proposition 3.5.**    (1) *The compositions* (3.1) *define the structure of an operad on the complexes $\{C_\bullet(\underline{\mathcal{M}}_{n+1})\}$.*

(2) *This operad structure induces the structure of an operad (in the category of graded vector spaces) on the homology* $\{H_\bullet(\underline{\mathcal{M}}_{n+1})\} = \{H_\bullet(\mathcal{M}_{n+1})\}$.

The chain operad is too large for our purposes: it consists of infinite dimensional spaces, and it makes little sense to describe algebras over it. The homology operad is too small: an algebra over it will always have vanishing differentials. The following theorem allows us to extract an intermediate operad, which turns out to be exactly the homotopy Lie operad.

**Theorem 3.6.** *There is a spectral sequence* $E^r_{p,q}$, $n-2 \leq p \leq 2n-4$, $-p \leq q \leq 0$, *possessing the following properties.*

(1) $E^r_{p,q} \Rightarrow H_{p+q}(\underline{\mathcal{M}}_{n+1}; k)$.

(2) $E^1_{p,q} = H_{p+q}(F_p, F_{p-1}; k)$, *where* $F_p$ *is the closure of the strata of dimension* $p$.

(3) *The complex*

$$(3.2) \qquad 0 \to E^1_{2n-4,0} \to \ldots \xrightarrow{d^1} E^1_{p,0} \xrightarrow{d^1} E^1_{p-1,0} \xrightarrow{d^1} \cdots \to E^1_{n-2,0} \to 0$$

*(the* $q = 0$th *row of the term* $E^1$, *with grading coming from the grading on chains from Section* 3.3, *but written with* $p$ *decreasing to the right) is naturally isomorphic to the* $n$th *component* $\mathcal{S}(n)$ *of the homotopy Lie operad. The operad structure on the* $E^1_{p,0}$'s *coming from* (3.1) *is compatible with the operad structure on the* $\mathcal{S}(n)$'s.

(4) *The spectral sequence collapses at the term* $E^2$, *i.e.,* $E^2 \cong E^\infty$.

(5) *The homology of the complex* (3.2) *is concentrated at the right end of the complex and is isomorphic to* $H_{n-2}(\underline{\mathcal{M}}_{n+1}; k) \cong H_{n-2}(\mathcal{M}_{n+1}; k) \cong \mathcal{L}(n)$ *(up to the signs of the action of the symmetric group, see Remark* 2.1*), where* $\mathcal{L}(n)$ *is the Lie operad.*

*Remark* 3.1. Deligne [10] studied the analogous spectral sequence associated with a general complex stratification and proved that it collapsed. The operad structure on the spectral sequence corresponding to the Deligne-Knudsen-Mumford compactification was noticed in [5]. The collapse of an analogous spectral sequence in the case of configuration space is due to Getzler and Jones [16].

*Proof.* The collapse of the spectral sequence follows from the collapse of the analogous spectral sequence for the configuration space, proved in [14, Proof of Theorem 4.4] and [16].

The assertion (5) about the cohomology of the top row of the spectral sequence follows from the collapse and the evaluation of the cohomology of the moduli space, which can be found in [33, 17].

To prove other assertions, let $F_p := \coprod_{\dim S^r_T \leq p} S^r_T$, $n-2 \leq p \leq 2n-4$, be the filtration of the space $\underline{\mathcal{M}}_{n+1}$ associated with all closed strata. This filtration generates a spectral sequence satisfying (1) and (2). Evidently,

$$F_p/F_{p-1} = \text{ the one-point compactification of } \coprod_{\substack{T \in \mathcal{T}(n)/\text{ or} \\ \dim S_T^r = -|T|-1 = p}} S_T^r$$

and for $p = n - 2$

$$F_{n-2} = \coprod_{\substack{T \in \mathcal{T}(n)/\text{ or} \\ \dim S_T^r = -|T|-1 = n-2}} S_T^r \cong \coprod_{\substack{T \in \mathcal{T}(n)/\text{ or} \\ |T| = 1-n}} (S^1)^{n-2},$$

$\coprod$ denoting a disjoint union. This gives us a calculation of the terms $E_{p,0}^1$. The calculation of the differential follows from item (6) of Theorem 3.4.  $\square$

**Proposition 3.7.** *For the spectral sequence of Theorem* 3.6*,*

$$E_{p,q}^1 = 0 \quad \text{for } q < 2-n \text{ and } q > 0,$$

*so that the term $E^1$, displayed in the third quadrant, looks like*

$$0 \to E_{2n-4,0}^1 \quad \to \ldots \to E_{p,0}^1 \quad \to E_{p-1,0}^1 \quad \to \ldots \to E_{n-2,0}^1 \quad \to 0$$
$$\vdots \qquad\qquad\qquad\qquad\qquad\qquad\qquad\qquad \vdots$$
$$0 \to E_{2n-4,2-n}^1 \to \ldots \to E_{p,2-n}^1 \to E_{p-1,2-n}^1 \to \ldots \to E_{n-2,2-n}^1 \to 0$$

*Proof.* Using Lefschetz duality, we have $E_{p,q}^1 = H_{p+q}(F_p, F_{p-1}) = H^{-q}(F_p \setminus F_{p-1})$. The Leray spectral sequence for the bundle

$$(S^1)^{2n-4-p} \to F_p \setminus F_{p-1} \xrightarrow{\pi} \overline{F}_{p+2-n} \setminus \overline{F}_{p+1-n},$$

where $\overline{F}_\bullet$ is the filtration for the Deligne-Knudsen-Mumford compactification indexed by the complex dimension, has the second term equal to $H^a(\overline{F}_{p+2-n} \setminus \overline{F}_{p+1-n}, H^b((S^1)^{2n-4-p}))$ and converges to $H^{a+b}(F_p \setminus F_{p-1})$. Notice that $0 \leq a \leq p+2-n$, and $0 \leq b \leq 2n-4-p$, where the former follows from the computation of the cohomology of the moduli spaces easily obtained from Arnold and F. Cohen's computation of the cohomology of configuration spaces. Therefore, for $-q = a + b > n - 2$, $E_{p,q}^1$ vanishes.  $\square$

**3.4. More decorations.** Let us define still larger spaces $\underline{\mathcal{N}}_n$, which will possess nicer properties than $\overline{\mathcal{M}}_n$ and $\underline{\mathcal{M}}_n$: they will make up an operad, like the $\overline{\mathcal{M}}_n$'s, and they will produce naturally the homotopy Lie operad $\mathcal{S}$, like the $\underline{\mathcal{M}}_n$'s.

The space $\underline{\mathcal{N}}_n$ is defined as the moduli space of stable $n$-punctured complex curves $C$ of genus 0 *decorated with relative phase parameters at double points and phase parameters at punctures*, which means that the sum of the arguments of germs of holomorphic coordinates on each irreducible component at each double point and the argument of a germ of holomorphic coordinate on each irreducible component at each puncture is chosen. Geometrically, this adds a choice of tangent direction at each puncture on each irreducible component of $C$ to the decoration of $C$ of Section 3.2. This is clearly a compactification of $\mathcal{N}_n$, the moduli space of nondegenerate decorated Riemann spheres .

Thus, the space $\underline{\mathcal{N}}_n$ is naturally a fibration over $\underline{\mathcal{M}}_n$ with fibre $(S^1)^n$. If we define a stratification of $\underline{\mathcal{N}}_n$ by pulling up the stratification of $\underline{\mathcal{M}}_n$, then the stratification will enjoy properties similar to those of Theorem 3.4, except that now the product $\mathcal{N}_{k+1} \times \mathcal{N}_{n_1+1} \times \cdots \times \mathcal{N}_{n_k+1}$ of spaces of nondegenerate decorated Riemann spheres will be fibred over the corresponding stratum with fibre $(S^1)^k$. This provides the spaces $\underline{\mathcal{N}}_{n+1}$, $n \geq 2$, with the structure of an operad. It is convenient to add the space $\underline{\mathcal{N}}_2$ defined as $(S^1)^2$ (the space of phase parameters $\theta_1$ and $\theta_2$ at 0 and $\infty$), defining the composition with an element $(\theta_1, \theta_2) \in \underline{\mathcal{N}}_2$ just by changing the phase parameter at the corresponding puncture by $\theta_1 + \theta_2$. This encodes the action of the rotation group on the spaces $\underline{\mathcal{N}}_{n+1}$. As in the previous section, the top row of the corresponding spectral sequence will be isomorphic to the tree operad $\mathcal{S}(n)$.

## 4. String theory

In this section, we show how the axioms of a CFT and hence a CSFT based on such a CFT provide an algebra over the homotopy Lie operad of Section 2 and hence the structure of a homotopy Lie algebra on the BRST complex of such a theory.

**4.1. Conformal field theory (CFT).** Conformal field theories we consider have central charge equal to 0 and are all at the tree level, i.e., the Riemann surfaces involved are only Riemann spheres.

Consider the Virasoro algebra Vir, which is the algebra of complex-valued vector fields on the circle in this text. Vir is generated by the elements $L_m = z^{m+1} \partial/\partial z$, $m \in \mathbb{Z}$, with the commutators given by the formula $[L_m, L_n] = (n-m)L_{m+n}$. By $V$ we will denote the complexification of this algebra $V := \text{Vir} \otimes_{\mathbb{R}} \mathbb{C} = \text{Vir} \oplus \overline{\text{Vir}}$.

Let $\mathcal{P}_n$ be the moduli space of nondegenerate Riemann spheres $\Sigma$ with $n$ punctures and holomorphic disks at each puncture (holomorphic embeddings of the standard disk $|z| < 1$ to $\Sigma$ centered at the puncture and not containing other punctures). The spaces $\mathcal{P}_{n+1}$, $n \geq 1$, form an operad under sewing Riemann spheres at punctures (cutting out the disks $|z| \leq r$ and $|w| \leq r$ for some $r = 1 - \epsilon$ at sewn punctures and identifying the annuli $r < |z| < 1/r$ and $r < |w| < 1/r$ via $w = 1/z$). The symmetric group interchanges punctures along with the holomorphic disks, as usual.

A *CFT* (*at the tree level*) consists of the following *data*:

(1) A topological vector space $\mathcal{H}$ (a *state space*).
(2) An action $T : V \otimes \mathcal{H} \to \mathcal{H}$ of the complexified Virasoro algebra $V$ on $\mathcal{H}$.
(3) A vector $|\Sigma\rangle \in \text{Hom}(\mathcal{H}^{\otimes n}, \mathcal{H})$ for each $\Sigma \in \mathcal{P}_{n+1}$ depending smoothly on $\Sigma$.

*Remark* 4.1. Physicists usually have a (possibly indefinite) Hilbert space as a state space $\mathcal{H}$. The structures we study do not involve any inner product on the state space, so we postulate it to be a vector space, although retaining the letter $\mathcal{H}$.

The natural extension of the action $T$ to an action of direct sums of $V$ on tensor powers of $\mathcal{H}$ will be denoted by the same letter: $T : V^{n+1} \otimes \operatorname{Hom}(\mathcal{H}^n, \mathcal{H}) \to \operatorname{Hom}(\mathcal{H}^n, \mathcal{H})$. We will use this convention later on, when we introduce more operators on $\mathcal{H}$. Here and henceforth we use the abbreviated notation: $\mathcal{H}^n := \mathcal{H}^{\otimes n}$.

To be honestly called a CFT, these data must satisfy the following compatibility *axioms*:

(4) $T(\mathbf{v})|\Sigma\rangle = |\delta(\mathbf{v})\Sigma\rangle$, where $\mathbf{v} = (v_1, \dots, v_{n+1}) \in V$ and $\delta$ is the natural action of $V^{n+1}$ on $\mathcal{P}_{n+1}$ by infinitesimal reparameterizations at punctures. In particular, $T(\mathbf{v})|\Sigma\rangle = T(\overline{\mathbf{v}})|\Sigma\rangle = 0$, whenever $\mathbf{v}$ can be extended to a holomorphic vector field on $\Sigma$ outside of the disks.

(5) The correspondence $\Sigma \mapsto |\Sigma\rangle$ defines the structure of an algebra over the operad $\mathcal{P}_{n+1}$ on the space of states $\mathcal{H}$.

*Remark* 4.2. Axiom (5) is equivalent to the factorization property or the sewing axiom of a CFT, see Segal [38], plus the so-called "manifest" symmetry property of the states $|\Sigma\rangle$ with respect to permutations of punctures. (As we have learned from Edward Witten, "'manifest' means nothing more than 'obvious' in plain English", cf. Zwiebach [49, footnote in Section 7.4]).

*Remark* 4.3. An equivalent definition of a CFT may be as follows: a CFT is an algebra over the operad $\mathcal{P}_{n+1}$. This would imply an action of $V^{n+1}$ on the states $|\Sigma\rangle$ via Axiom (4).

**4.2. String background.** We will now consider string backgrounds, or in other words, CFT's with ghosts and with total central charge $c = 0$, although our results can be easily generalized to generalized string backgrounds or topological CFT's, gadgets which do not contain the operators $c$ below.

A *string background* (*at the tree level*) is a CFT based on a vector space $\mathcal{H}$ with the following additional *data*:

(1) A $\mathbb{Z}$-grading $\mathcal{H} = \bigoplus_{i \in \mathbb{Z}} \mathcal{H}_i$ on the state space.
(2) An action of the Clifford algebra $C(V \oplus V^*)$, which is denoted usually by $b : V \otimes \mathcal{H} \to \mathcal{H}$ and $c : V^* \otimes \mathcal{H} \to \mathcal{H}$ for generators of the Clifford algebra, the degree of $b$ is $-1$, and the degree of $c$ is 1.
(3) A differential $Q : \mathcal{H} \to \mathcal{H}$, $Q^2 = 0$, of degree 1, called a *BRST operator*.

The graded space $\mathcal{H}$ with the operator $Q$ is called a *BRST complex*. For $\psi \in \mathcal{H}_i$, the degree $\operatorname{gh} \psi := i$ is called the *ghost number*.

*Remark* 4.4. Usually, the space $\mathcal{H}$ of a string background is constructed from a CFT based on a space of "matter" $\mathcal{H}_\mathrm{m}$ and a "ghost" CFT based on a space $\mathcal{H}_\mathrm{gh}$ as the tensor product $\mathcal{H} = \mathcal{H}_\mathrm{m} \otimes \mathcal{H}_\mathrm{gh}$, the grading coming from the second factor. In that case, the CFT's $\mathcal{H}_\mathrm{m}$ and $\mathcal{H}_\mathrm{gh}$ must be more general than the ones we consider, because $\mathcal{H}_\mathrm{m}$ and $\mathcal{H}_\mathrm{gh}$ have nontrivial central charges. But for the resulting string background, the central charge can be made 0 by an appropriate choice of $\mathcal{H}_\mathrm{m}$.

These data must satisfy the following *axioms*:

(4) $[T(v_1), b(v_2)] = b([v_1, v_2])$ and $[T(v_1), c(v_2^*)] = c((\mathrm{ad}^* v_1) v_2^*)$, as in any BRST complex over $V$ or $Vir$.
(5) $[Q, T(v)] = 0$, $\{Q, b(v)\} = T(v)$, $\{Q, c(v^*)\} = c(dv^*)$, where $dv^* \in \Lambda^2(V)^*$ is the Lie algebra differential of the 1-cochain $v^* \in V^*$, as in any BRST complex.
(6) $\mathrm{gh}\,|\Sigma\rangle = 0$.
(7) $b(\mathbf{v})|\Sigma\rangle = b(\overline{\mathbf{v}})|\Sigma\rangle = 0$ for any $\mathbf{v} \in V^{n+1}$ extended holomorphically on $\Sigma$.
(8) $Q|\Sigma\rangle = 0$.

*Remark* 4.5. Following A. S. Schwarz's idea, one can define a more general string background in the same fashion as we defined a CFT in Remark 4.3: a generalized string background is an algebra over the operad $\mathcal{P}_{n+1}^{sr}$ of semirigid $N = 2$ super Riemann surfaces introduced by Distler and Nelson [12]. The operators $b(v)$ and $Q$ on states $|\Sigma\rangle$ can be read off as infinitesimal transformations of the super structure on a semirigid surface $\Sigma$, similar to the action $T(v)$ of the Virasoro in the usual CFT case. The connection to semirigid supergravity of Distler and Nelson has been pointed out in Horava's paper [20].

**4.3. A morphism of complexes.** One of the nicest implications of a string background is the construction of a morphism of complexes $\mathcal{H}^{\otimes n} \to \Omega^\bullet(\mathcal{P}_{n+1})$, $n \geq 1$, from the tensor power of the BRST complex $\mathcal{H}$ to the de Rham complex of the space $\mathcal{P}_{n+1}$. We will use a somewhat partially dual picture and construct for each $n$ a $\mathrm{Hom}(\mathcal{H}^n, \mathcal{H})$-valued form $\Omega_{n+1} = \sum_{r \geq 0} \Omega_{n+1}^r$, $\deg \Omega_{n+1}^r = r$, on the space $\mathcal{P}_{n+1}$:

$$(4.1) \qquad \Omega_{n+1}(\mathbf{v}_1, \ldots, \mathbf{v}_r) := \Omega_{n+1}^r(\mathbf{v}_1, \ldots, \mathbf{v}_r) := b(\tilde{\mathbf{v}}_1) \ldots b(\tilde{\mathbf{v}}_r)|\Sigma\rangle,$$

where $\mathbf{v}_i$, $1 \leq i \leq r$, is a tangent vector to the space $\mathcal{P}_{n+1}$ at the point $\Sigma$ and $\tilde{\mathbf{v}}_i$ is its pullback to an element of $V^{n+1}$, acting by infinitesimal reparameterizations at punctures.

**Theorem 4.1.**

$$(d - Q)\Omega_{n+1} = 0$$

*Remark* 4.6. The only structure we needed from a string background for our purposes was this collection of forms $\Omega_{n+1}$. One can show that a collection of such forms together with a natural operad condition is equivalent to a generalized string background, see Segal [39] or Getzler [14].

*Proof.* More precisely, fixing $n$ and omitting the subscript $n + 1$, we have to prove that for each $r \geq 0$

$$(4.2) \qquad\qquad d\Omega^{r-1} = Q\Omega^r,$$

where $\Omega^{-1} = 0$ by definition. Let us use induction on $r$. For $r = 0$, (4.2) is $0 = Q|\Sigma\rangle$, which has been postulated. Suppose that Equation (4.2) is true for some $r \geq 0$. To make the induction step, it suffices to prove that for any tangent vector $\mathbf{v}$ to $\mathcal{P}_{n+1}$,

$$\iota(\mathbf{v}) d\Omega^r = \iota(\mathbf{v}) Q \Omega^{r+1},$$

where $\iota$ is the contraction of a differential form with a tangent vector. From (4.1), it is clear that for all $r \geq 0$

$$(4.3) \qquad \iota(\mathbf{v})\Omega^r = b(\tilde{\mathbf{v}})\Omega^{r-1}.$$

According to Axiom (4) of Section 4.1, we have that the Lie derivative $\mathcal{L}(\mathbf{v}) := \{d, \iota(\mathbf{v})\}$ of a form along a tangent vector $\mathbf{v}$ is equal to $T(\tilde{\mathbf{v}})$:

$$(4.4) \qquad \{d, \iota(\mathbf{v})\} = \mathcal{L}(\mathbf{v}) = T(\tilde{\mathbf{v}}) = \{Q, b(\tilde{\mathbf{v}})\}.$$

Since the operator $d$ acts geometrically and $b$ acts in coefficients, we have

$$[d, b(\tilde{\mathbf{v}})] = 0.$$

Similarly,

$$[Q, \iota(\mathbf{v})] = 0.$$

Using these equations, we obtain

$$\begin{aligned}
\iota(\mathbf{v}) d\Omega^r &= \mathcal{L}(\mathbf{v})\Omega^r - d\iota(\mathbf{v})\Omega^r \\
&= T(\tilde{\mathbf{v}})\Omega^r - db(\tilde{\mathbf{v}})\Omega^{r-1} \\
&= T(\tilde{\mathbf{v}})\Omega^r - b(\tilde{\mathbf{v}})d\Omega^{r-1} \\
&= T(\tilde{\mathbf{v}})\Omega^r - b(\tilde{\mathbf{v}})Q\Omega^r \\
&= Q\iota(\mathbf{v})\Omega^{r+1} \\
&= \iota(\mathbf{v})Q\Omega^{r+1}. \quad \square
\end{aligned}$$

Some crucial properties of these differential forms are their equivariance under the operad map and the actions of the permutation group.

**Theorem 4.2.**

$$\gamma^*\Omega_{n+1} = \gamma(\Omega_{k+1}; \Omega_{n_1+1}, \dots, \Omega_{n_k+1})$$

*where $n = n_1 + \cdots + n_k$, $\gamma^* : \Omega^\bullet(\mathcal{P}_{n+1}) \to \Omega^\bullet(\mathcal{P}_{k+1} \times \mathcal{P}_{n_1+1} \times \cdots \mathcal{P}_{n_k+1})$ is the pullback of the operad map $\mathcal{P}_{k+1} \times \mathcal{P}_{n_1+1} \times \cdots \times \mathcal{P}_{n_k+1} \to \mathcal{P}_n$ and the other $\gamma$ is the operad map in the endomorphism operad. Similarly, for all $\sigma$ in $\Sigma_n$, the permutation group, we have*

$$\sigma^*\Omega_{n+1} = \Omega_{n+1} \circ \sigma.$$

*Proof.* Both are proved by using the axioms of a string background.  $\square$

*Remark* 4.7. Theorems 4.1 and 4.2 insure that the maps $C_\bullet(\mathcal{P}_{n+1}) \to \operatorname{Hom}(\mathcal{H}^n, \mathcal{H})$ given by $C \mapsto \int_C \Omega_{n+1}$ makes $\mathcal{H}$ into an algebra over $\{C_\bullet(\mathcal{P}_{n+1})\}$ which, in turn, makes absolute BRST cohomology, $H^\bullet$, into an algebra over the operad $\{H_\bullet(\mathcal{P}_{n+1})\}$.

*Remark* 4.8. Dually, the above properties of the forms $\Omega_{n+1}$ along with their behavior with respect to the identity of the operad $\{\mathcal{P}_{n+1}\}$ can be formulated as follows. If the $\Omega_{n+1}$'s were regarded as maps from $\operatorname{Hom}(\mathcal{H}^n, \mathcal{H})^* \to \Omega^\bullet(\mathcal{P}_{n+1})$, then they would define the structure of a coalgebra over the DG cooperad of differential forms on $\mathcal{P}_{n+1}$ on the complex $\mathcal{H}$.

**4.4. Closed string-field theory (CSFT).** Suppose that we have a string background as above. Then a *closed string-field theory* over this background consists of a choice of smooth mappings $s = s_{n+1} : \underline{\mathcal{N}}_{n+1} \to \mathcal{P}_{n+1}$ for each $n \geq 1$. The images $s(\underline{\mathcal{N}}_{n+1}) \subset \mathcal{P}_{n+1}$ of these mappings are called *string vertices*. The string vertices must satisfy the following *axiom*, which basically governs two things: the string vertices must be closed under sewing and symmetric with respect to permutations of punctures.

> The collection of mappings $s = s_{n+1} : \underline{\mathcal{N}}_{n+1} \to \mathcal{P}_{n+1}$, $n \geq 1$, defines a morphism of operads.

These mappings may be constructed as homotopy inverses of the natural mappings $\mathcal{P}_{n+1} \to \mathcal{N}_{n+1} \hookrightarrow \underline{\mathcal{N}}_{n+1}$, which are homotopy equivalences. More exactly, the mapping $s_{n+1} : \underline{\mathcal{N}}_{n+1} \to \mathcal{P}_{n+1}$ is composed by pushing $\underline{\mathcal{N}}_{n+1}$ away from the boundary in $\mathcal{N}_{n+1}$, so that it forms the complement to a tubular neighborhood of the boundary, and then lifting it to $\mathcal{P}_{n+1}$ along the homotopy equivalence $\mathcal{P}_{n+1} \to \mathcal{N}_{n+1}$.

Furthermore, the fact that the mappings $s_{n+1} : \mathcal{N}_{n+1} \to \mathcal{P}_{n+1}$ are invariant with respect to composition with elements of $\mathcal{N}_2$ and $\mathcal{P}_2$ implies that the $s_{n+1}$'s are equivariant under the $U(1)^{n+1}$ action corresponding to the rotation of the phases of the coordinates, *i.e.* if $\phi_j$ is the phase parameterizing rotations about the $j$-th puncture of a point $\Sigma$ in $\mathcal{N}_{n+1}$ and $\theta_j$ is the phase parameterizing rotations about the $j$th puncture of $s(\Sigma)$ in $\mathcal{P}_{n+1}$ then $s_* \partial/\partial\phi_j = \partial/\partial\theta_j$. As a consequence, a rotation in the relative phase at the $j$-th double point of

$$\Sigma' = \gamma(\Sigma; \Sigma_1, \ldots, \Sigma_k)$$

in $\underline{\mathcal{N}}_{n+1}$ (which corresponds to rotation of the phase at the initial point of $\Sigma_j$) maps under $s = s_{n+1}$ to the change in

$$s(\Sigma') = \gamma(s(\Sigma); s(\Sigma_1), \ldots, s(\Sigma_k))$$

induced by rotation of the initial disk in $s(\Sigma_j)$.

*Remark* 4.9. The string vertices defined here are a bit different from Zwiebach's string vertices: the latter are the images of mappings of $\underline{\mathcal{M}}_{n+1}$ to an analog of the space $\mathcal{P}_{n+1}$ with forgotten phases at punctures, rather than of $\underline{\mathcal{N}}_{n+1}$ to $\mathcal{P}_{n+1}$. Nevertheless, we can use his construction with minimal area metrics to check the existence theorem. This is a very nontrivial result.

**Theorem 4.3.** *String vertices exist.*

While the idea of the construction of string vertices is due to Zwiebach, [49], a mathematically rigorous account can be found in Wolf and Zwiebach [47]. The advantage of our approach of axiomatizing the string vertices is that we allow a certain freedom to the choice of vertices. Physically speaking, we deal with an arbitary solution of the string equation.

The construction of Wolf and Zwiebach assigns to Riemann spheres with punctures a holomorphic disk centered about each puncture defined *up to a rigid rotation or multiplication by a phase.* If, in addition, a tangent direction were specified at each puncture then there would be a canonical chart centered about each puncture resulting in the map $s : \mathcal{N}_{n+1} \to \mathcal{P}_{n+1}$. They do so by assigning to each Riemann sphere $\Sigma$ with punctures $p_1, \cdots, p_{n+1}$ a complete metric which solves a minimal area problem for an admissible class of metrics. An admissible metric is a one which satisfies the condition that the length of any nontrivial homotopy closed curve is greater than or equal to $2\pi$. This minimal area metric has nice properties. First of all, about every puncture $p_i$ is a neighborhood which is isometric to a semi-infinite flat cylinder of circumference $2\pi$ – this cylinder is naturally foliated by closed geodesics of circumference $2\pi$. The cylinder must end somewhere – let $\mathcal{C}_i(0)$ be this closed geodesic of length $2\pi$ where the cylinder comes to an end. Let $\mathcal{C}_i(r)$ where $r \geq 0$ be the closed geodesic of length $2\pi$ on the semiinfinite cylinder a distance $r$ from $\mathcal{C}_i(0)$. Fix a number $l > \pi$ once and for all. The curve $\mathcal{C}_i(l)$, called a *coordinate curve*, bounds a semi-infinite cylinder isometric to $S^1 \times \mathbb{R}_{\geq l}$ (endowed with standard metrics so that $S^1$ has circumference $2\pi$) thereby inducing natural coordinates $(\sigma_i, \tau_i)$ on $S^1 \times \mathbb{R}_{\geq l}$ such that $\sigma_i \in [0, 2\pi)$ and $\tau_i \geq l$. Curves of constant $\tau_i$ correspond to $\mathcal{C}_i(\tau_i)$. Furthermore, we assume that the geodesic $\sigma_i = 0$ lies along the tangent direction at $p_i$ thereby assigning a marked point to the coordinate curve. Defining $z_i = e^{l - \tau_i + i\sigma_i}$, we have assigned a unit disk in $\mathbb{C}$ to the neighborhood of $p_i$ which defines a map $s : \mathcal{N}_{n+1} \to \mathcal{P}_{n+1}$. This map allows us to compose the $i$th puncture (for any $i = 1, \cdots, n$) of $\Sigma$ in $\mathcal{N}_n$ to the $(n'+1)$st puncture of $\Sigma'$ in $\mathcal{N}_{n'}$ by sewing together their images in $\mathcal{P}_{n+1}$ and $\mathcal{P}_{n'+1}$, respectively, to obtain a point in $\mathcal{P}_{n+n'-1}$ and then projecting it to $\mathcal{N}_{n+n'-1}$. This composition is associative precisely because sewing together $\Sigma$ and $\Sigma'$ with their minimal area metrics results in a Riemann sphere with a minimal area metric. These compositions make $\mathcal{N}_n$ into an operad. It is clear that in performing the previous composition of $\Sigma$ and $\Sigma'$ along the $i$th puncture, we would have obtained the same resulting Riemann sphere had we rotated the tangent directions by the same amount on the $(n+1)$th puncture on $\Sigma$ and the $i$th puncture on $\Sigma'$. Since only the relative phase between the two punctures matter when performing the sewing, $s$ extends to a morphism of operads $s : \underline{\mathcal{N}}_{n+1} \to \mathcal{P}_{n+1}$.

Using the string vertices, we can consider the restrictions $s_{n+1}^* \Omega_{n+1}$ of the forms $\Omega_{n+1}$ on $\mathcal{P}_{n+1}$ to $\underline{\mathcal{N}}_{n+1}$. From now on, we will deal with these restrictions often and will still denote them by $\Omega_{n+1}$.

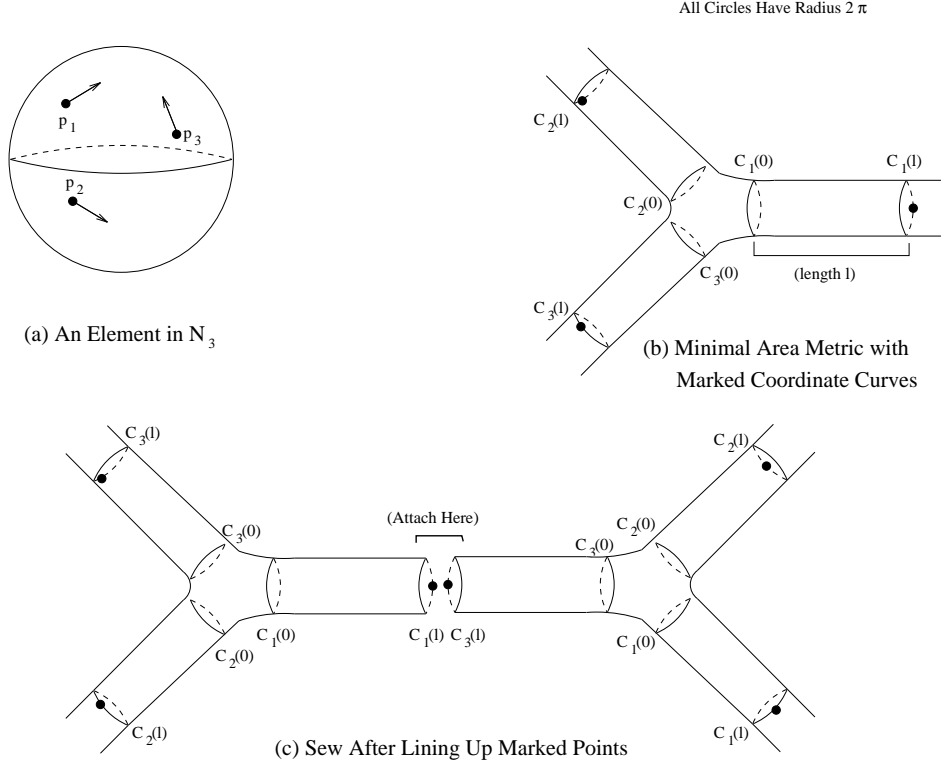

(a) An Element in N$_3$

(b) Minimal Area Metric with Marked Coordinate Curves

(c) Sew After Lining Up Marked Points

FIGURE 1. Minimal Area Metrics and Sewing

### 4.5. Algebra over an operad.
To get a homotopy Lie algebra structure for a given CSFT, we will first obtain an algebra over the operad of chains in $\underline{\mathcal{M}}_{n+1}$ of Section 3.3, and then restrict the structure to the smaller operad $\mathcal{S}(n)$. Since CSFT data apply to the operad $\underline{\mathcal{N}}_{n+1}$ rather than $\underline{\mathcal{M}}_{n+1}$, we have to descend to $\underline{\mathcal{M}}_{n+1}$. That can be done by passing to the following relative BRST complex (which is called semirelative in the physical literature).

Let $\mathcal{H}_{\mathrm{rel}}$ be the subspace annihilated by $b_0^- := b(L_0^-)$ and $T_0^- := T(L_0^-)$, where $L_0^- := L_0 - \overline{L}_0 = \frac{1}{2\pi i} \partial / \partial \theta \in V$, $z = e^{2\pi i \theta}$, is the generator of rigid rotations of the circle. In terms of the BRST complex, this is the space of BRST cochains relative to $\mathfrak{u}(1)$, which is the one-dimensional subalgebra $\langle L_0^- \rangle$ of rigid rotations. The differential $Q$ induces a differential here; we have a subcomplex.

It would be natural to restrict values of our $\mathrm{Hom}(\mathcal{H}^n, \mathcal{H})$-valued form $\Omega_{n+1}$ to the subspace $(\mathcal{H}_{\mathrm{rel}})^n \subset \mathcal{H}^n$. Nothing tells us that the result will be contained in $\mathcal{H}_{\mathrm{rel}}$, but we can use the natural projection onto the quotient space $\mathcal{H}^{\mathrm{rel}} := \mathcal{H}/(b_0^- \mathcal{H} + T_0^- \mathcal{H})$ of relative $\mathfrak{u}(1)$-chains. Fortunately, $\mathcal{H}^{\mathrm{rel}}$ is canonically isomorphic to $\mathcal{H}_{\mathrm{rel}}$. The isomorphism is defined by the formula:

$$(4.5) \qquad \mathcal{H}^{\mathrm{rel}} \to \mathcal{H}_{\mathrm{rel}},$$
$$[x] \mapsto 2\pi i b_0^- x_0,$$

where $x_0$ is the $T_0^-$-invariant (i.e., rotationally invariant) component of a vector $x \in \mathcal{H}$. The image lies indeed in the subspace $\mathcal{H}_{\mathrm{rel}}$, because $(b_0^-)^2 = 0$ and $[T_0^-, b_0^-] = 0$. The mapping is an isomorphism, because it has an inverse

$$\mathcal{H}_{\mathrm{rel}} \to \mathcal{H}^{\mathrm{rel}},$$
$$x \mapsto \frac{1}{2\pi i}[c_0^- x],$$

where $c_0^- := c((L_0^-)^*)$, $(L_0^-)^* \in V^*$ being the corresponding element of the dual basis.

*Remark* 4.10. To make the above considerations valid in the context of generalized string backgrounds, where the operator $c_0^-$ does not exist, we should just postulate that (4.5) is an isomorphism.

Thus, restricting our form $\Omega_{n+1}$ to $(\mathcal{H}_{\mathrm{rel}})^n$ and then mapping the result to $\mathcal{H}_{\mathrm{rel}}$ via (4.5), we obtain a $\mathrm{Hom}((\mathcal{H}_{\mathrm{rel}})^n, \mathcal{H}_{\mathrm{rel}})$-valued form $\Omega'_{n+1}$ on $\underline{\mathcal{N}}_{n+1}$. It turns out that this form is basic, i.e., is pulled up from some form $\omega_{n+1}$ on $\underline{\mathcal{M}}_{n+1}$.

**Proposition 4.4.**     (1) *Each* $\mathrm{Hom}((\mathcal{H}_{\mathrm{rel}})^n, \mathcal{H}_{\mathrm{rel}})$-*valued form* $\Omega'_{n+1}$ *is basic with respect to the natural projection* $p : \underline{\mathcal{N}}_{n+1} \to \underline{\mathcal{M}}_{n+1}$, *i.e.,* $\Omega'_{n+1} = p^*\omega_{n+1}$ *for some form* $\omega_{n+1}$ *on* $\underline{\mathcal{M}}_{n+1}$.
(2) *There holds the following formula:*

$$(4.6) \qquad \omega_{n+1}(\mathbf{v}_1, \ldots, \mathbf{v}_r) = \omega_{n+1}^r(\mathbf{v}_1, \ldots, \mathbf{v}_r) = 2\pi i b_0^-(b(\tilde{\mathbf{v}}_1) \ldots b(\tilde{\mathbf{v}}_r)|\Sigma\rangle)_0,$$

*where the operator* $b_0^-$ *acts in* $\mathrm{Hom}$ *as usual, for* $1 \le j \le r$, $\mathbf{v}_j$ *is a tangent vector to the space* $\mathcal{P}_{n+1}$ *at the point* $\Sigma$ *and* $\tilde{\mathbf{v}}_j$ *is its pullback to an element of* $V^{n+1}$.

*Proof.*     (1) To verify that the form $\Omega'_{n+1}$ is basic, we have to show that it is annihilated by the operators $\iota(\mathbf{v})$ and $\mathcal{L}(\mathbf{v})$, where $\mathbf{v} = (0, \ldots, 0, \partial/\partial\theta_j, 0, \ldots, 0)$ is the tangent vector at $\Sigma \in \mathcal{P}_{n+1}$ generated by rotations at the $j$th puncture, $j = 1, \ldots, n+1$. From Equation (4.3), we see that $\iota(\mathbf{v})\Omega'_{n+1} = 0$. Equation (4.4) implies $\mathcal{L}(\mathbf{v})\Omega'_{n+1} = 0$.
(2) Obvious by construction from $\Omega$ and the nature of the relativization process.  $\square$

The $\mathrm{Hom}((\mathcal{H}_{\mathrm{rel}})^n, \mathcal{H}_{\mathrm{rel}})$-valued forms $\omega_{n+1}$ on $\underline{\mathcal{M}}_{n+1}$ are just what we need, as we can see from the following theorem.

**Theorem 4.5.** *The correspondence*

$$(4.7) \qquad C_\bullet(\underline{\mathcal{M}}_{n+1}) \to \mathrm{Hom}((\mathcal{H}_{\mathrm{rel}})^n, \mathcal{H}_{\mathrm{rel}}),$$

$$C \mapsto \int_C \omega_{n+1},$$

*defines the structure of an algebra over the operad $C_\bullet(\underline{\mathcal{M}}_{n+1})$ on the space $\mathcal{H}_{\mathrm{rel}}$.*

*Proof.* Stokes' theorem along with Theorem 4.1 implies that the mapping (4.7) commutes with differentials. The morphism (4.7) preserves the degree, because $\mathrm{gh}\,|\Sigma\rangle = 0$, therefore $\mathrm{gh}\,\omega_{n+1}^r = -r - 1$, and that was defined to be the degree of a chain $C$ of dimension $r$ in Section 3.3.

The equivariance with respect to the symmetric group is evident from the definition of its action. Thus, we need only prove that (4.7) respects the operad composition, i.e., to prove a factorization property of the forms $\omega_{n+1}$:

$$\int_{\gamma(C;C_1,\dots,C_k)} \omega_{n+1} = \gamma(\int_C \omega_{k+1}; \int_{C_1} \omega_{n_1+1}, \dots, \int_{C_k} \omega_{n_k+1}),$$

$n = n_1 + \cdots + n_k$, which will follow from

$$(4.8) \qquad \oint \omega_{n+1} = \gamma(\omega_{k+1}; \omega_{n_1+1}, \dots, \omega_{n_k+1}),$$

where $\oint$ denotes integration along the fibres $(S^1)^k$ of the natural projection

$$p : \overline{S}_T^r \to \underline{\mathcal{M}}_{k+1} \times \underline{\mathcal{M}}_{n_1+1} \times \cdots \times \underline{\mathcal{M}}_{n_k+1}$$

and $T = \gamma(\delta_k; \delta_{n_1}, \dots, \delta_{n_k})$, see Theorem 3.4.

We know that both sides of Equation (4.8) are forms on $\underline{\mathcal{N}}_{k+1} \times \underline{\mathcal{N}}_{n_1+1} \times \cdots \times \underline{\mathcal{N}}_{n_k+1}$ which were pulled back from $\mathcal{P}_{k+1} \times \mathcal{P}_{n_1+1} \times \cdots \times \mathcal{P}_{n_k+1}$. From now on, let us use the same notation $\underline{\mathcal{N}}_{j+1}$ for the image of the space $\underline{\mathcal{N}}_{j+1}$ in $\mathcal{P}_{j+1}$. We need to show that the values taken by these forms at each $r$-tuple $(\mathbf{v}_1, \dots, \mathbf{v}_r)$ of tangent vectors at a point $(\Sigma; \Sigma_1, \dots, \Sigma_k) \in \underline{\mathcal{N}}_{k+1} \times \underline{\mathcal{N}}_{n_1+1} \times \cdots \times \underline{\mathcal{N}}_{n_k+1}$ are equal. Indeed, by (4.3), we have

$$\iota(\mathbf{v}_1) \dots \iota(\mathbf{v}_r) \oint \omega_{n+1}^{k+r} = b(\mathbf{v}_1) \dots b(\mathbf{v}_r) \oint \omega_{n+1}^k$$

and

$$\iota(\mathbf{v}_1) \dots \iota(\mathbf{v}_r) \gamma(\omega_{k+1}; \omega_{n_1+1}, \dots, \omega_{n_k+1})^r$$
$$= b(\mathbf{v}_1) \dots b(\mathbf{v}_r) \gamma(\omega_{k+1}; \omega_{n_1+1}, \dots, \omega_{n_k+1})^0$$
$$= b(\mathbf{v}_1) \dots b(\mathbf{v}_r) \gamma(2\pi i b_0^- |\Sigma\rangle_0; 2\pi i b_0^- |\Sigma_1\rangle_0, \dots, 2\pi i b_0^- |\Sigma_k\rangle_0),$$

where the superscript means the "component of the corresponding degree" of a differential form. Thus, it remains to show that

$$(4.9) \qquad \oint \omega_{n+1}^k = \gamma(2\pi i b_0^- |\Sigma\rangle_0; 2\pi i b_0^- |\Sigma_1\rangle_0, \dots, 2\pi i b_0^- |\Sigma_k\rangle_0),$$

where the left-hand side is evaluated at the point $\Sigma \times \Sigma_1 \times \cdots \times \Sigma_k$.

In order to effect the integration in (4.9), let us introduce coordinates $\theta_j$, $j = 1, \ldots, k$, in a fibre $(S^1)^k$, where $\theta_j$ is the phase parameter for the initial puncture on $\Sigma_j$, which is glued with the $j$th puncture on $\Sigma$. Let

$$\Sigma' = \gamma(\Sigma; \Sigma_1, \ldots, \Sigma_k)$$

be a point in $\underline{\mathcal{N}}_{n+1}$ at which the form $\omega_{n+1}$ is evaluated in (4.9). By construction of $\Omega$ and subsequent definition of $\omega$, we have on each fibre

$$\omega_{n+1}^k = 2\pi i b_0^- b(\partial/\partial\theta_1) \ldots b(\partial/\partial\theta_k)|\Sigma'\rangle_0 d\theta_1 \ldots d\theta_k.$$

Since a CFT and a CSFT define a morphism of operads $\underline{\mathcal{N}}_{\bullet+1} \to \mathcal{E}nd\,(\mathcal{H})$, we have the corresponding equation for states:

$$|\Sigma'\rangle = \gamma(|\Sigma\rangle; |\Sigma_1\rangle, \ldots, |\Sigma_k\rangle).$$

Applying $b(\partial/\partial\theta_j)$ to this equation and integrating over the circle in the fibre $(S^1)^k$ parameterized by $\theta_j$, we get

$$\int_0^1 b(\partial/\partial\theta_j)|\Sigma'\rangle d\theta_j = \int_0^1 \gamma(|\Sigma\rangle; |\Sigma_1\rangle, \ldots, 2\pi i b_0^- |\Sigma_j\rangle, \ldots, |\Sigma_k\rangle) d\theta_j$$
$$= \gamma(|\Sigma\rangle; |\Sigma_1\rangle, \ldots, 2\pi i b_0^- |\Sigma_j\rangle_0, \ldots, |\Sigma_k\rangle).$$

The second equality holds because full integration of a function on the circle produces the rotationally invariant average value, i.e., the $T_0^-$-invariant component. Iterating this procedure for all $j = 1, \ldots, k$ and applying the mapping $2\pi i b_0^-()_0$, see (4.5), we obtain the factorization equation (4.9).  □

**Corollary 4.6.** *The correspondence*

$$H_p(F_p, F_{p-1}) \to \mathrm{Hom}((\mathcal{H}_{\mathrm{rel}})^n, \mathcal{H}_{\mathrm{rel}}),$$
$$Z \mapsto \int_Z \omega_{n+1},$$

*defines the structure of an algebra over the homotopy Lie operad $\mathcal{S}(n)$ on the relative state space $\mathcal{H}_{\mathrm{rel}}$.*

*Proof.* This is an obvious corollary of Theorem 4.5. For the upper row of the spectral sequence is a suboperad of the chain operad: indeed, there is a natural mapping

$$H_p(F_p, F_{p-1}) \to C_p(\underline{\mathcal{M}}_{n+1}),$$
$$\text{the fundamental class } F_p \mapsto \text{the chain } F_p,$$

which defines a morphism of operads.  □

This corollary can be reformulated as the following result of Zwiebach [49].

**Corollary 4.7.** *A CSFT defines the structure of a homotopy Lie algebra on the space* $\mathcal{H}_{\mathrm{rel}}$ *of relative states. The brackets defining this structure are given by the formula:*

$$[\cdot, \ldots, \cdot] = \int_{\mathcal{M}_{n+1}} \omega_{n+1} = \int_{\mathcal{M}_{n+1}} \omega_{n+1}^{2n-4} \in \mathrm{Hom}((\mathcal{H}_{\mathrm{rel}})^n, \mathcal{H}_{\mathrm{rel}}).$$

*Proof.* A homotopy Lie algebra structure on $\mathcal{H}_{\mathrm{rel}}$ is yielded by Corollary 4.6 via Theorem 2.4. Also according to Theorem 2.4, the $n$-ary bracket is given by the corolla $\delta_n$, which corresponds to the fundamental cycle $\mathcal{M}_{n+1} \in H_{2n-4}(\underline{\mathcal{M}}_{n+1}, \partial \underline{\mathcal{M}}_{n+1})$ due to Theorem 3.6. $\square$

**4.6. Topological gravity.** We obtained the homotopy Lie structure using the action of the huge chain operad $C_\bullet(\underline{\mathcal{M}}_{n+1})$ and restricting the structure to a suboperad, which was isomorphic to the homotopy Lie operad $\mathcal{S}$. Now we will study what happens at the (co)homology level. Let $H_{\mathrm{rel}}^\bullet$ denote the relative BRST cohomology, i.e., the cohomology of the operator $Q$ on the space $\mathcal{H}_{\mathrm{rel}}$ of relative states.

**Corollary 4.8.** *The correspondence*

$$H_p(\underline{\mathcal{M}}_{n+1}) \to \mathrm{Hom}((H_{\mathrm{rel}}^\bullet)^n, H_{\mathrm{rel}}^\bullet),$$

$$Z \mapsto \int_Z \omega_{n+1},$$

*defines the structure of an algebra over the operad* $H_\bullet(\mathcal{M}_{n+1}) = H_\bullet(\underline{\mathcal{M}}_{n+1})$ *on the relative BRST cohomology* $H_{\mathrm{rel}}^\bullet$.

*Proof.* This is a general fact: if we have a morphism of operads in the category of complexes, it induces a morphism of operads on the cohomology. $\square$

**Definition 4.1.** A *topological gravity* (at the tree level) is an algebra over the operad $H_\bullet(\underline{\mathcal{M}}_{n+1})$ of homology of the real compactification of the moduli spaces.

**Corollary 4.9.** *CSFT data defines a topological gravity based on the space* $H_{\mathrm{rel}}^\bullet$ *of relative BRST cohomology.*

*Remark* 4.11. A topological gravity can be defined alternatively as an algebra over the homology operad $H_\bullet(\overline{\mathcal{M}}_{n+1})$ of the Deligne-Knudsen-Mumford compactification of the moduli space. This version fits the context of intersection theory on the moduli space better, but the connection of this theory with string theory is not that obvious. It would be interesting to describe algebras over $H_\bullet(\overline{\mathcal{M}}_{n+1})$ algebraically, as a collection of operations, generators and identities, in the spirit of Getzler's description [15] of algebras over the operad $H_\bullet(\mathcal{M}_{n+1})$, which we are going to use in the following corollary.

**Corollary 4.10.** *A CSFT defines the structure of a* gravity algebra *on the space* $H^\bullet_{\text{rel}}$ *of relative BRST cohomology, that is, a collection of brackets* $\{x_1, \ldots, x_n\}$, $x_1, \ldots, x_n \in H^\bullet_{\text{rel}}$, *of degree* $-1$ *satisfying the following relations*

$$\sum_{1 \le i < j \le k} \epsilon(i,j) \{\{x_i, x_j\}, x_1, \ldots, \hat{x}_i, \ldots, \hat{x}_j, \ldots, x_k, y_1, \ldots, y_l\}$$
$$= \{\{x_1, \ldots, x_k\}, y_1, \ldots, y_l\},$$

*where* $k \ge 2$, $l \ge 0$ *and the right-hand side is interpreted as zero if* $l = 0$. *The sign* $\epsilon(i,j)$ *is the sign picked up by rearranging the sequence* $x_i, x_j, x_1, \ldots, \hat{x}_i, \ldots, \hat{x}_j, \ldots,$ $x_k$ *to the sequence* $x_1, \ldots, x_k$ *in commutative superalgebra. The brackets defining this structure are given by the formula:*

$$\{\cdot, \ldots, \cdot\} = \int_{a\ point\ \Sigma \in \mathcal{M}_{n+1}} \omega_{n+1} = 2\pi i b_0^- |\Sigma\rangle_0 \in \text{Hom}((H^\bullet_{\text{rel}})^n, H^\bullet_{\text{rel}}).$$

**4.7. The Batalin-Vilkovisky (BV) algebra.** This algebraic structure differs from the previous algebraic structures in two ways. First of all, it exists for any string background and does not depend upon the choice of a particular closed string-field theory. In fact, the results of this section can be generalized to more general so-called topological conformal field theories. Secondly, it is defined on the absolute BRST cohomology $H^\bullet$ rather than the relative BRST cohomology.

**Definition 4.2.** A *Gerstenhaber algebra* is a graded commutative and associative algebra $A$ together with a bracket $[\cdot, \cdot] : A \otimes A \to A$ of degree $-1$, such that for all homogeneous elements $x$, $y$, and $z$ in $A$,

$$[x, y] := -(-1)^{(|x|-1)(|y|-1)} [y, x],$$
$$[x, [y, z]] = [[x, y], z] + (-1)^{(|x|-1)(|y|-1)} [y, [x, z]],$$

and

$$[x, yz] = [x, y]z + (-1)^{|x|(|y|-1)} y[x, z].$$

If, in addition, $A$ has an operation $\Delta : A \to A$ of degree $-1$ such that $\Delta^2 = 0$ and

$$[x, y] = \Delta(xy) - ((\Delta x)y + (-1)^{|x|} x(\Delta y)),$$

then $A$ is said to be a *Batalin-Vilkovisky (BV) algebra*.

Associated to the Gerstenhaber and BV algebras are certain special operads. Let $D$ be the unit disk in the complex plane and let $\mathcal{F}(n)$, $n \ge 1$, be the space of all maps $f$ from $\coprod_{i=1}^n D \to D$ such that $f$, when restricted to each disk, is the composition of translation and multiplication by an element of $\mathbb{C}^\times$ and images of $f$ are disjoint. $\{\mathcal{F}(n)\}$ forms an operad by composition in the natural way and is called the *framed little disks operad*. This operad has a suboperad $\{\mathcal{D}(n)\} \overset{i}{\hookrightarrow} \{\mathcal{F}(n)\}$ called the *little disks operad*, where $\mathcal{D}(n)$ consists of those maps in $\mathcal{F}(n)$ which, when restricted to each disk, are the compositions of translations and multiplications by positive real numbers. By general arguments, the homologies of these topological operads are operads. F. Cohen [9] proved that the category of algebras over the operad $\{H_\bullet(\mathcal{D}(n))\}$ was isomorphic to the category of Gerstenhaber algebras. Similarly, Getzler [14] proved that the category of algebras over the operad $\{H_\bullet(\mathcal{F}(n))\}$ was isomorphic to the category of BV algebras.

These operads naturally appear in the context of a string background. Consider a map $j : \mathcal{F}(n) \hookrightarrow \mathcal{P}_{n+1}$ which is defined as follows. Let $z$ be a standard coordinate on $\mathbb{CP}^1$. Given an element of $\mathcal{F}(n)$, identify the complex plane in which the large disk sits with the image of $z$ then we get $\mathbb{CP}^1$ with $n$ embedded holomorphic disks. In addition, let the $(n+1)$st disk be the complement of the large disk $D$, a holomorphic disk around $z = \infty$. This gives a point in $\mathcal{P}_{n+1}$ and completes the definition of the map $j$. Notice that $j$ is a morphism of operads, which induces a morphism between $\{H_\bullet(\mathcal{F}(n))\}$ and $\{H_\bullet(\mathcal{P}_{n+1})\}$. Furthermore, $j$ is a homotopy equivalence for $n \geq 2$. Therefore, $j$ induces an isomorphism of operads between $\{H_\bullet(\mathcal{F}(n))\}$ and $\{H_\bullet(\mathcal{P}_{n+1})\}$, except for $n = 1$ where it is an embedding. From the composition $\mathcal{D}(n) \overset{i}{\hookrightarrow} \mathcal{F}(n) \overset{j}{\to} \mathcal{P}_{n+1}$, we obtain another morphism of operads $\{H_\bullet(\mathcal{D}(n))\} \to \{H_\bullet(\mathcal{P}_{n+1})\}$. This leads to the following theorem.

**Theorem 4.11.** *Given a string background, the absolute BRST cohomology $H^\bullet$ is a vector space graded by ghost number, which admits the structure of a BV algebra. It is convenient to describe the operations in the BV algebra as being induced from operations defined on BRST cochains. The associative product is induced from*

$$x \cdot y := |\Sigma\rangle(x \otimes y) \quad \forall x, y \in \mathcal{H},$$

*where $\Sigma$ is a point ( and, hence, a 0-cycle ) in $\mathcal{P}_3$. The bracket is induced from*

$$[x, y] := (-1)^{|x|}(\int_C \Omega_3^1)(x \otimes y) \quad \forall x, y \in \mathcal{H},$$

*where $|x|$ is the ghost degree of $x$ and $C$ is a cycle in $\mathcal{P}_3$ whose class is the image of the generator in $H_1(\mathcal{D}(2)) \cong k$ given by the full counterclockwise rotation of one disk about the other. Finally, $\Delta$ is induced from*

$$\Delta x := (\int_h \Omega_2^1)(x) \quad \forall x \in \mathcal{H},$$

*where $h$ is a cycle in $\mathcal{P}_2$ whose class is the image of the generator of $H_1(\mathcal{F}(1)) \cong k$, the full rotation of the disk $D$ in the counterclockwise direction. Moreover,*

$$\Delta x := 2\pi i b_0^- x_0, \quad \forall x \in \mathcal{H},$$

*where the subscript $0$ means the $T_0^-$-invariant part of the vector.*

*Proof.* Given a string background, the integration of the differential forms $\Omega_{n+1}$ on $\mathcal{P}_{n+1}$ with values in $\text{Hom}(\mathcal{H}^n, \mathcal{H})$ over chains in $\mathcal{P}_{n+1}$ gives $\mathcal{H}$ the structure of an algebra over the operad of chains on $\mathcal{P}_{n+1}$. This induces the structure at the level of the (co)homologies, that is, the absolute BRST cohomology $H^\bullet$ forms an algebra over the operad $\{H_\bullet(\mathcal{P}_{n+1})\}$. But the latter is isomorphic as an operad to the homology of the framed little disks operad $\{H_\bullet(\mathcal{F}(n))\}$. Therefore, by Getzler's theorem [14], the absolute BRST cohomology, $H^\bullet$, forms a BV algebra. To obtain the correct homology classes which give rise to our basic operations, we need only to make some simple observations.

Ignoring $\Delta$ for a moment, $H^\bullet$ has the structure of a Gerstenhaber algebra induced by the action of the operad $\{H_\bullet(\mathcal{D}(n))\}$ on $H^\bullet$ through the map induced by the composition $\mathcal{D}(n) \overset{i}{\hookrightarrow} \mathcal{F}(n) \overset{j}{\to} \mathcal{P}_{n+1}$. The dot product is a binary operation of ghost degree 0 which arises from the image of a generator in $H_0(\mathcal{D}(2)) \cong k$ in $H_0(\mathcal{P}_3)$ via the map induced by $j \circ i$. More directly, the dot product arises from the class of a $\Sigma$ in $H_0(\mathcal{P}_3)$ where $\Sigma$ is a point in $\mathcal{P}_3$. In that case, the dot product is induced by $\int_\Sigma \Omega_3 = |\Sigma\rangle$. Similarly, the bracket is a binary operation of ghost degree $-1$ arising from the image of a generator in $H_1(\mathcal{D}(2)) \cong k$ in $H_1(\mathcal{P}_3)$, call the image $C$, which can be described as above. The bracket is then defined by integration of $\Omega_3^1$ over $C$ with an insertion of the annoying factor of $(-1)^{|x|}$ which is necessary to insure that $[x, y] = -(-1)^{(|x|-1)(|y|-1)}[y, x]$. (Without this factor, we would have $[x, y] = (-1)^{|x||y|}[y, x]$ from the equivariance of the operad action under the permutation group.) Finally, $\Delta$ is a unary operation of ghost degree $-1$ arising from the image of a generator $H_1(\mathcal{F}(1)) \cong k$ in $H_1(\mathcal{P}_2)$ denoted by $h$ which is described above. The other expression for $\Delta$ is a straightforward exercise in the application of the axioms of a string background. $\square$

*Remark* 4.12. Although different choices of the cycles representing the same homology classes above induce the same operations of a BV algebra on BRST cohomology, the corresponding operations on the BRST cochains, $\mathcal{H}$, will certainly depend upon these choices. In this way, we can work at the level of BRST cochains, as is done in Lian-Zuckerman [29], where the operations would be given by particular cycles $\Sigma$, $C$, and $h$ in $\mathcal{D}(n)$, $\mathcal{F}(n)$ and $\mathcal{P}_{n+1}$ for $n = 1, 2$ and where the relations satisfied by these operations would be obtained from the action of the operad of chains of $\{\mathcal{P}_{n+1}\}$ upon $\mathcal{H}$.

Also, taking the $T_0^-$-invariant part of all of the differential forms in a string background is still a string background. Furthermore, the operations induced on absolute BRST cohomology are the same as before since $[Q, b_0^-] = T_0^-$ implies that the only nontrivial BRST cocycles are in the kernel of $T_0^-$ and, therefore, the only nontrivial operations on BRST cohomology are induced from the component with zero $T_0^-$. Nonetheless, at the level of BRST cochains, one would obtain different operations.

*Remark* 4.13. In the case of a meromorphic string background where the vector space $\mathcal{H}$ is a *topological vertex operator algebra* (TVOA) (or, rather, some completion thereof), the results of Huang [21] may allow for the construction of similar forms $\Omega$, which are holomorphic. In his case, the BV algebra structure on BRST cohomology is precisely the one discovered by Lian-Zuckerman [29]. Explicit expressions for the bracket and the dot product may be written in terms of the elements of the TVOA and a direct comparision with the formulas of Lian-Zuckerman is possible. Huang's construction can be used to obtain smooth forms, as well, by wedging his holomorphic forms with their antiholomorphic counterparts which are associated to the isomorphic vector space of the opposite chirality.

*Remark* 4.14. Another remarkable algebraic aspect of BV theory, which remains beyond the scope of this paper, is its connection with odd symplectic structures. We refer the interested reader to the papers [14, 19, 35, 36].

**4.8. Concluding remarks.** To obtain the homotopy Lie structure, we actually used only the upper row of the spectral sequence $E^1$ of Theorem 3.6. At the same time, the whole $E^1$ operad, which is more tractable than the entire chain operad $C_\bullet(\underline{\mathcal{M}}_{n+1})$, carries much more information than its upper row. The "on-shell" part of the structure it gives, i.e., the part related to the homology $H_\bullet(\underline{\mathcal{M}}_{n+1})$ of the operad $E^1$, is the structure of Getzler's gravity algebra, as we have just seen. On the other hand, another row of $E^1$ gives the so-called commutative homotopy algebras, which we anticipate to play a special role in meromorphic string theory (TVOA's). Finally, we hope to describe algebras over the whole $E^1$ operad as homotopy Gerstenhaber algebras, homotopy analogues of the Gerstenhaber algebras studied by Lian and Zuckerman [29]. The role of these algebras in string theory is unclear at the moment.

From the point of view of moduli spaces and CFT, it would be more natural to consider several initial punctures (inputs) as well as several outputs, instead of separating one of them as an input and all others as outputs. Corresponding generalizations of operads are known as PROP's in topology. The moduli space PROP that we would deal with in the scope of this paper is in fact equivalent to the moduli space operad, and apparently, at the moment there is no need to complicate the situation on the algebraic side with operations from $\mathcal{H}^m$ to $\mathcal{H}^n$. But as soon as $m$-ary operations are well-understood, a consistent theory of $(m, n)$-ary operations would be at least interesting.

The larger and more interesting piece of moduli spaces for higher genera and higher perturbations of string theory, correspondingly, remain beyond the scope of this work. The matter is that punctured Riemann surfaces of higher genera form an object which is slightly more general than an operad: apart from the operad compositions, corresponding to sewing (or attaching) Riemann surfaces, one should consider sewings of a Riemann surface with itself, which forms a new handle. The corresponding generalization of vertex operator algebras was considered by Zhu [48], but the corresponding algebraic structures which have appeared in string theory, see Verlinde [45] and Zwiebach [49], are yet to be understood.

From the topological point of view, we have related the homology of moduli spaces to a homotopy Lie algebra structure on the state space. In comparison to that, we should mention an interesting recent work of M. Betz and R. L. Cohen [6] relating the topology of moduli spaces to the Steenrod algebra.

*Acknowledgment.* This work is partly a result of our collaboration with Huang at the seminar organized by Gerstenhaber and Stasheff at the University of Pennsylvania in 1992/93. We greatly benefited from weekly interactions with Murray Gerstenhaber and Yi-Zhi Huang and are very grateful to them. We would also like to thank P. Deligne, E. Getzler, V. Hinich, P. Horava, M. Kapranov, A. Losev, P. May, J. Morava, V. Schechtman, A. S. Schwarz, G. Segal, G. Zuckerman and B. Zwiebach for helpful discussions. The second author would like to thank the University of Pennsylvania for hospitality during his leave and Lehigh University for a subsequent visiting appointment. The third author is grateful to Princeton University, where major part of the work has been accomplished, and RIMS at Kyoto University, where he found an excellent opportunity to finish this paper.

University of Pennsylvania

University of North Carolina

University of Pennsylvania